\documentclass[aps,prl,superscriptaddress]{revtex4}
\usepackage{times}
\usepackage{float}
\usepackage[normalem]{ulem}
\usepackage{epsfig}
\usepackage{soul,xcolor}
\usepackage{epstopdf}
\usepackage{amsmath,amssymb,amscd}
\usepackage{graphicx}
\usepackage{lipsum}
\usepackage{dcolumn}
\usepackage{bm}
\usepackage{ifsym}
\usepackage{braket}
\usepackage{amssymb}
\usepackage{graphicx}

\usepackage{blindtext}
\usepackage{dashrule}
\usepackage{array}
\usepackage{tikz}
\usepackage{textcomp}
\usepackage{verbatim}
\usepackage{pdfpages}
\usetikzlibrary{trees,decorations.pathmorphing}

\begin{document}

\title{Collective  Energy Transfer to a Spectator Atom via Multi-Center Intermolecular Coulombic Decay  }

\author{ Saroj Barik}
\affiliation{Department of Physics, Indian Institute of 
Technology Madras, Chennai, India}

\author{ Pratikkumar Thakkar}
\affiliation{Department of Physics, Indian Institute of 
Technology Madras, Chennai, India}

\author{Siddhartha S. Payra}
\affiliation{Department of Physics, Indian Institute of 
Technology Madras, Chennai, India}

\author{Yash Lenka}
\affiliation{Department of Physics, Indian Institute of 
Technology Madras, Chennai, India}

\author{ Y. Sajeev}
\email{sajeevy@barc.gov.in} 

\affiliation{Theoretical Chemistry Section, Bhabha 
Atomic Research Centre, Mumbai 400 085, India}

\author{G. Aravind}
\email{garavind@iitm.ac.in} 
\affiliation{Department of Physics, Indian Institute of 
Technology Madras, Chennai, India}


\begin{abstract} { Molecular mechanisms that enable collective and upconverted energy transfer from multiple photoacceptors to a non-absorbing spectator reaction center are highly desirable for efficient light-energy utilization.  Here, we show that intermolecular Coulombic decay (ICD), a nonlocal energy-relaxation channel in photoexcited molecules, offers an avenue for such a novel energy-transfer mechanism. On irradiation of pyridine-argon gas mixture at 266 nm and at low laser intensities, we observed a surprisingly dominant formation of argon cations.  Measurements of the laser‐power dependence, together with systematic studies of Ar$^+$ yield versus laser intensity and molecular density, reveal that ICD mediates the collective funneling of excitation energy from multiple photoexcited pyridine molecules to a non-photoabsorbing argon atom, leading to its ionization. The density of the reaction center offers an efficient handle to optimize this collective  energy transfer.  This mechanism opens new avenues in light-harvesting design and may help explain the remarkable resistance of biomolecules to photodamage.}  \end{abstract}

\maketitle

\begin{center}
    \textbf{Published in Phys. Rev. Lett. \copyright 2026 American Physical Society. 
 } \\
    \textbf{Saroj \textit{et al.}, Phys. Rev. Lett. 136, 093201 (2026)}
 \url{DOI: https://doi.org/10.1103/9gqh-9yq6}
\end{center}
\pagebreak
\maketitle


Intermolecular Coulombic decay (ICD) is one of the most efficient and ubiquitous non-radiative relaxation pathways by which electronically excited molecules in a molecular environment collectively relax through the ionization of neighboring species~\cite{cederbaum_1997, Kuleff_2010, Demekhin_2013, Jahnke_2020, Jahnke_2015}. Since its theoretical prediction~\cite{cederbaum_1997}, ICD has been experimentally observed in a variety of molecular environments, including rare gas clusters, liquid water, hydrogen-bonded dimers, and $\pi$-stacked aromatic systems~\cite{Jahnke_2020, Ren_2022, Zheng22, Zhou2022, Jahnke_2010, Dominic_2013, LaForge_2014, Kumagai_2018, Richter_2018, Nagaya_2016, Dubrouil_2015, Nicolas_2010, Kirill_2014, Havermeir_2010, Cabrera_2020, LaForge_2019} Kuleff \textit{et al.} studied ICD in multiply-excited clusters \cite{Kuleff_2010, stimulated}. However, most of these studies have focused on the ICD relaxation originating from the inner-valence ionized states involving pairwise intermolecular interactions. The strong dependence of ICD rates on intermolecular interactions initially limited observations to weakly bound clusters; however, recent advances have extended its applicability to more complex and even unbound molecular systems ~\cite{Jahnke_2020, Ren_2022, Zheng22, Zhou2022, Jahnke_2010, Dominic_2013, LaForge_2014, Kumagai_2018, barik2022, barik2023, icdrev}.

A recent breakthrough demonstrated that ICD can also occur following $\pi$-$\pi^*$ photoexcitation under ambient UV light, where multiple $\pi$ conjugated molecules collectively relax via a highly efficient multi center ICD process~\cite{barik2022, barik2023}. This finding establishes photoinduced multi center ICD as a broadly relevant mechanism in molecular photochemistry and highlights its potential role in nonlocal redistribution and dissipation of electronic excitation energy. In contrast to conventional two-center ICD, this mechanism involves the collective relaxation of multiple excited molecules, enabling the release of excess energy through the ionization of one of them. 

Building on this observation, a fundamental question emerges: Can multiple photoexcited molecules collectively funnel their energy to a non-absorbing spectator unit? Such an upconverted energy transfer mechanism would represent a significant extension of ICD beyond donor-only systems and could provide a compelling explanation for energetically demanding processes observed in astrophysical and biochemical environments--where the energy of ambient photons falls well below the threshold required to directly activate non-absorbing reaction centers.
Addressing this question is the central objective of the present study, which investigates the feasibility of collective energy transfer to a non-absorbing spectator reaction center in a model system relevant to light-harvesting mechanisms.

To address this question in a molecular environment, we employ a gas-phase model system in which photoabsorption and energy reception are clearly separated. Specifically, we investigate photoabsorption in an unbound mixture of pyridine monomers diluted in argon (Ar) gas~\cite{NOTES}. Electronically unbound pyridine molecules serve as photoabsorbing units, while argon acts as a non-absorbing spectator reaction center. 
Photoionization of the non-photoabsorbing Ar reaction center  occurs via multi-center ICD, with Ar$^+$ detection serving as a direct signature of this collective upconverted energy transfer. The strong absorption of pyridine at 266 nm, contrasted with argon’s transparency at this wavelength, renders the system ideal for probing this process.

 \color{black}

Collective energy transfer from multiple photoexcited pyridine molecules to an argon atom requires close spatial proximity of the interaction partners. At sufficiently short distances, excited pyridine molecules can transiently associate while simultaneously interacting with a nearby argon atom, thereby creating favorable conditions for ICD. An interaction region characterized by frequent molecular encounters is therefore expected to facilitate such proximity and enable collective energy transfer.

In our experiment, this condition is achieved using a pulsed supersonic jet that generates a locally high density of pyridine molecules and argon atoms in the laser-interaction region of the time-of-flight (TOF) mass spectrometer. The jet is directly injected, without a skimmer, between the first two TOF electrodes, creating a collision-rich environment. Similar collisional conditions have been shown to promote efficient associative interactions between multiple excited molecular species~\cite{barik2022}. Accordingly, in the present experiment, such collisions increase the likelihood that excited pyridine molecules and argon atoms approach sufficiently closely for collective energy transfer. Further technical details are provided in the Supplemental Material~\cite{NOTES}.

The pyridine-argon mixture, expanded between the electrodes as described above, was then irradiated with a 266 nm photon beam. Even at low laser intensities between $10^4-10^6$ Wcm$^{-2}$, we observed 
a surprisingly dominant formation of argon cations, as shown in Figure 
\ref{massspec}. All the cations that were observed in the 
undiluted pyridine experiments \cite{barik2022} were also 
observed.

\begin{figure}[h]
\centering
\includegraphics[width=1\columnwidth]{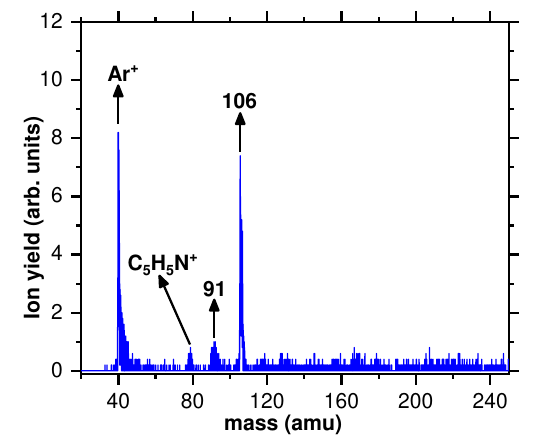}
\caption{\textbf{TOF mass spectrum of cations:} A supersonic expansion of a pyridine-argon mixture, prepared by diluting pyridine vapor with 1 atm of Ar, was irradiated with 266 nm laser light at an intensity of 1 $\times$ 10$^6$ Wcm$^{-2}$. The resulting TOF mass spectrum shows dominant formation of  Ar$^+$ and pyridine-derived cations. Raw data were smoothed using adjacent averaging in Origin 7.5.  }  

\label{massspec}
\end{figure}

\begin{figure}
    \centering
    \includegraphics[width=1\columnwidth]{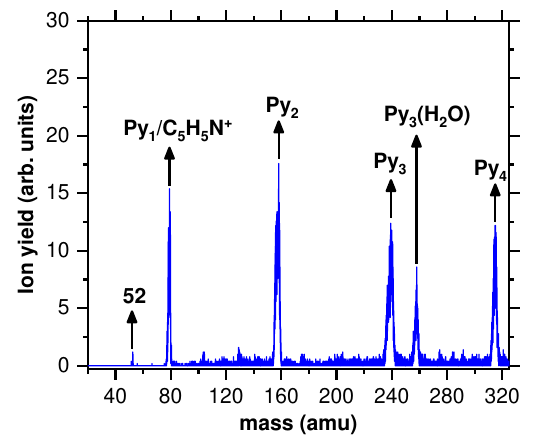}
    \caption{\textbf{Cation TOF spectrum from skimmed supersonic expansion:} A pyridine-argon mixture (2 atm) is expanded through a 1mm skimmer and irradiated with 266 nm photons.  The spectrum shows cations of pyridine, its fragments, and clusters, with no formation of Ar$^+$. Heavier-than-pyridine cations reported by Barik et al. \cite{barik2022} are also absent due to the collision-free nature of the skimmed beam.}
    \label{skim}
\end{figure}

In the present experiments, it is important to rule out the formation of Ar$^+$ via multiphoton absorption in Ar and via the impact of photoelectrons or ions from pyridine, which are accelerated in the TOF electric field. We could simultaneously rule out the role of the above mentioned processes with an important test experiment, as follows. We selected the central part of pyridine-argon supersonic beam using a skimmer of 1 mm aperture before letting the gas beam into the interaction chamber. We maintained the conditions of the laser beam and the TOF electric field to be the same as those used in the experiments without a skimmer. Irradiation of this skimmed gas mixture at 266 nm resulted in the formation of pyridine monomer and cluster cations; but, most notably, it did not produce Ar$^+$, as shown in Figure~\ref{skim}.

A skimmed supersonic beam being essentially collision-free rendered the close proximity of the ICD participants improbable. In fact, on irradiation of the skimmed pyridine-Ar gas mixture, the heavier-than-pyridine cations which were observed by Barik \textit{et al.} \cite{barik2022} to be formed under photoassociation-assisted ICD were largely suppressed. This collision-free nature of the skimmed beam also lead to the observation of intact cluster cations, which were not observed in the unskimmed beam experiments due to the collisional fragmentation of the neutral clusters even before they could reach the interaction spot. Barik \textit{et al.} \cite{barik2022} had also confirmed the absence of neutral clusters at the collision-prone interaction region with their electron-impact experiments. In this test experiment, wherein the laser and TOF electric field conditions were the same as that for the experiments without skimmer, the absence of Ar$^+$ formation not only emphasizes the critical requirement of close proximity for the ICD participants but also unambiguously rules out both the MPI and electron/ion -impact ionization of Ar in the present experiments. 
In fact, at the laser intensities employed here, we do not observe Ar$^+$ when pure argon gas is irradiated at 266 nm (see \cite{NOTES} for details). There are no accessible resonant intermediate states in argon at one-, two-, or three-photon energies that could facilitate MPI at this wavelength. Further experiments that rule out the formation of Ar$^+$ ions via electron/ion impact are discussed in the Supplemental Material~\cite{NOTES}.

With the skimmed-beam experiment ruling out both the MPI of Ar at our laser intensities and the electron/ion-impact ionization of argon under our TOF electric field conditions, we now examine the origin of Ar$^+$ formation under the standard (unskimmed) conditions. The photon energy at 266 nm (4.661 eV) is well below the ionization threshold of argon (15.75 eV), requiring the absorption of at least four photons to ionize a single atom. Interestingly, a log-log plot of the yield of Ar$^+$ versus laser intensity shows a slope of 1, far below the value of 4 expected for a four-photon MPI process.

The results from the skimmed versus unskimmed beam experiments, the observation of Ar$^+$ upon irradiation of the pyridine-argon mixture but not of the pure-argon beam under our laser intensities and TOF electric-field conditions, and the linear laser-power dependence collectively indicate that a mechanism other than direct photon absorption by argon is operative. This rules out any significant contribution from multiphoton ionization of argon. These findings set the stage for identifying collective energy transfer via multi-center ICD as the underlying process. The linear dependence of the yield of Ar$^+$ with the laser power is consistent with a single-photon absorption event in pyridine followed by nonlocal energy transfer to argon. This behavior aligns with ICD mechanisms, where the laser-power dependence reflects the initial photoexcitation step~\cite{Kuleff_2010, LaForge_2014, muller}. These findings confirm that Ar$^+$ cations are produced exclusively via a Coulombic decay pathway involving the collective transfer of electronic energy from multiple $\pi^*$-excited pyridine molecules to a single Ar atom.  

We now turn to the molecular mechanism underlying Coulombic decay processes in photoexcited pyridine molecules that lead to Ar$^+$ formation under our experimental conditions. In their $\pi^*$-excited states, pyridine molecules exhibit long-range associative covalent interactions, enabling the transient formation of electronically bound dimers even in the absence of ground-state binding~\cite{barik2022}. As a result, dominant ICD channels are expected to originate from excited-state dimers formed via associative collisions under our collision-prone experimental conditions. The total energy deposited by two photons in these dimers lies only marginally above the ionization threshold of pyridine, and intramolecular vibrational energy redistribution (IVR) is likely to render this energy off-resonant for direct ionization~\cite{barik2022}. Under such conditions, relaxation via three or more than three centre 
 ICD pathway emerges as the most efficient decay channel,
particularly in a collision-rich environment \cite{barik2022}.

These doubly excited-state pyridine dimers may provide a plausible pathway for ICD through interactions with additional excited monomers or dimers, leading to transient three- or four-center ICD configurations. Under our collision-rich experimental conditions, such multi-center configurations are most plausibly accessed through sequential two-body encounters, rather than concerted many-body collisions, due to the higher probability of successive collisions. While the present experiment does not directly resolve such intermediate complexes, this scenario is consistent with prior studies demonstrating efficient multi-center ICD in electronically coupled excited units~\cite{barik2022}. A quantitative estimate supporting the feasibility of this sequential pathway is provided in the Supplemental Material~\cite{NOTES}.

Two factors are expected to favor the likelihood of such sequential ICD events. 
First, the effective collision cross section for encounters involving excited-state dimers is increased as a result of their larger spatial extent upon dimerization. 
Second, the likelihood of these encounters depends on the lifetime of the excited electronic states. 
Pyridine exhibits both a picosecond-lived S$_1$ state and a longer-lived triplet state accessed via intersystem crossing, with $\pi$-$\pi^*$ character~\cite{TRIPLET}. 
The presence of this longer-lived state can increase the temporal window for collision-mediated ICD, even if the branching ratio into the triplet channel is modest.

We now analyze what happens when an argon atom is introduced into the photoexcited pyridine gas. The insertion of argon gives rise to a remarkable and previously unexplored ICD channel involving the argon atom and two excited pyridine dimers. 
While the present measurements do not uniquely determine the microscopic geometry of the interacting species, the magnitude of the Ar$^+$ signal is consistent with ICD involving argon interacting with more than one photoexcited pyridine unit.

\begin{figure}[h]
\centering
\includegraphics[scale=0.75]{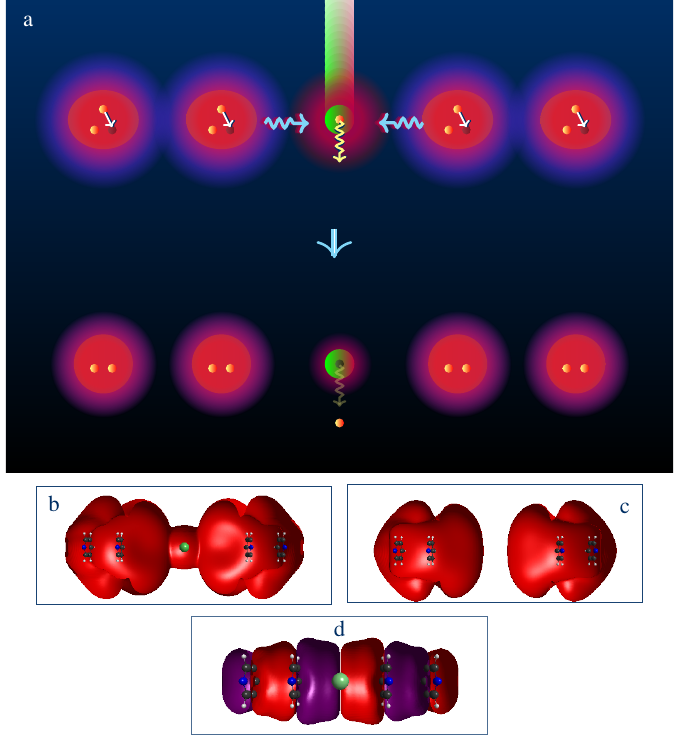}\\
\vskip 0pt
\caption{\textbf{Schematic and theoretical illustration of a collision-mediated ICD scenario involving argon and photoexcited pyridine dimers.}\
\textbf{a},  Green: argon atom; red: pyridine molecules. Valence electron density is shown as a haze, with the outer blue haze representing photoexcited electron density.   The figure illustrates a plausible configuration in which a diffusing argon atom enters the region between two electronically excited pyridine dimers, a scenario that can facilitate multi-center interactions leading to ICD. \textbf{b-c}, Comparison of total electron density surfaces (iso-density: $1\times 10^{-6}$) with and without Ar illustrates wavefunction delocalization induced by argon diffusion.  \textbf{d}, Intermolecular orbital formed via covalent mixing of the pyridine 1$\pi$ orbitals and the argon  $p_z$ orbital,  which provides a plausible theoretical pathway for energy transfer associated with argon ionization via ICD. }
\label{mech}
\end{figure}

  To rationalize how such interactions may occur and lead to Ar ionization under our experimental conditions, we consider the most probable collision scenarios. The rapid thermal diffusion of argon atoms into transient interstitial regions between otherwise non-interacting electronically excited pyridine dimers provides a plausible route for multi-center ICD-mediated argon ionization. This scenario is further supported by quantum chemical calculations on representative argon-pyridine cluster geometries, which indicate—beyond electrostatic contributions-a viable ionization pathway arising from multi-center interactions enabled by covalent mixing of pyridine $\pi$ orbitals with the argon $p_z$ orbital (see Figure~\ref{mech}). Such a covalent contribution is consistent with earlier reports of heavier-than-parent cations observed in ICD experiments on pyridine~\cite{barik2022}. Notably, this interaction persists even when the argon atom is located more than $10$~\AA~ from the excited molecules, suggesting an extended spatial range.  While fluorescence can compete at large intermolecular separations \cite{Kuleff_2010}, the rapid diffusion of Ar atoms under our collision-prone, room-temperature conditions leads to  short-range encounters, enabling ICD to dominate and be efficiently observed.

       From this perspective, the process may be viewed as a sequence of two-body scattering events, in which a relatively extended excited dimer assembly serves as a composite target for a diffusing argon atom (see Figure 4). The large spatial extent of such assemblies enhances the effective interaction volume and thus the probability of ICD-relevant encounters. Taken together, these results provide qualitative theoretical support for a plausible molecular mechanism rather than a definitive quantum-dynamical description. A quantitative estimate of the ICD-event yield leading to Ar ionization within this collisional scenario is provided in the Supplemental Material~\cite{NOTES}.

One of the key experimental observations supporting this molecular mechanism is the linear dependence of the Ar$^+$ ion yield on laser intensity (see also the discussion of Figure~\ref{Ar_rel}  below). Under the present laser conditions, Ar is not directly photoexcited; therefore, the photon-dependent rate-limiting step is the initial single-photon excitation of individual pyridine monomers. At low photon flux, where the formation of noninteracting dimer-dimer units is limited by the density of excited molecules rather than by laser intensity, the number of such units becomes linearly proportional to the excitation rate. Subsequent processes-such as dimerization, the formation of noninteracting dimer-dimer units, and the entry of a fast-moving argon atom into the shared interaction region of these units proceed efficiently, enabling Ar ionization via collective energy pooling. As a result, the Ar$^+$ yield exhibits a linear intensity dependence (slope $\sim$ 1), consistent with the experiments of La Forge \textit{et al.} \cite{LaForge_2014}, where collective autoionization of multiply excited He atoms produced a similar slope.

It must be emphasized that, under our collisional conditions, the term \textit{dimer} refers generally  to a  configuration of two interacting monomers, and not necessarily to a fully stabilized dimer. The term \textit{non-interacting dimer-dimer units} refers to transient pairs of such excited-state dimers that do not initially interact with each other but are capable of initiating intermolecular energy transfer upon the diffusion of an argon atom into their interstitial region.

\begin{figure} [h]
\centering  
\includegraphics[scale=0.345]{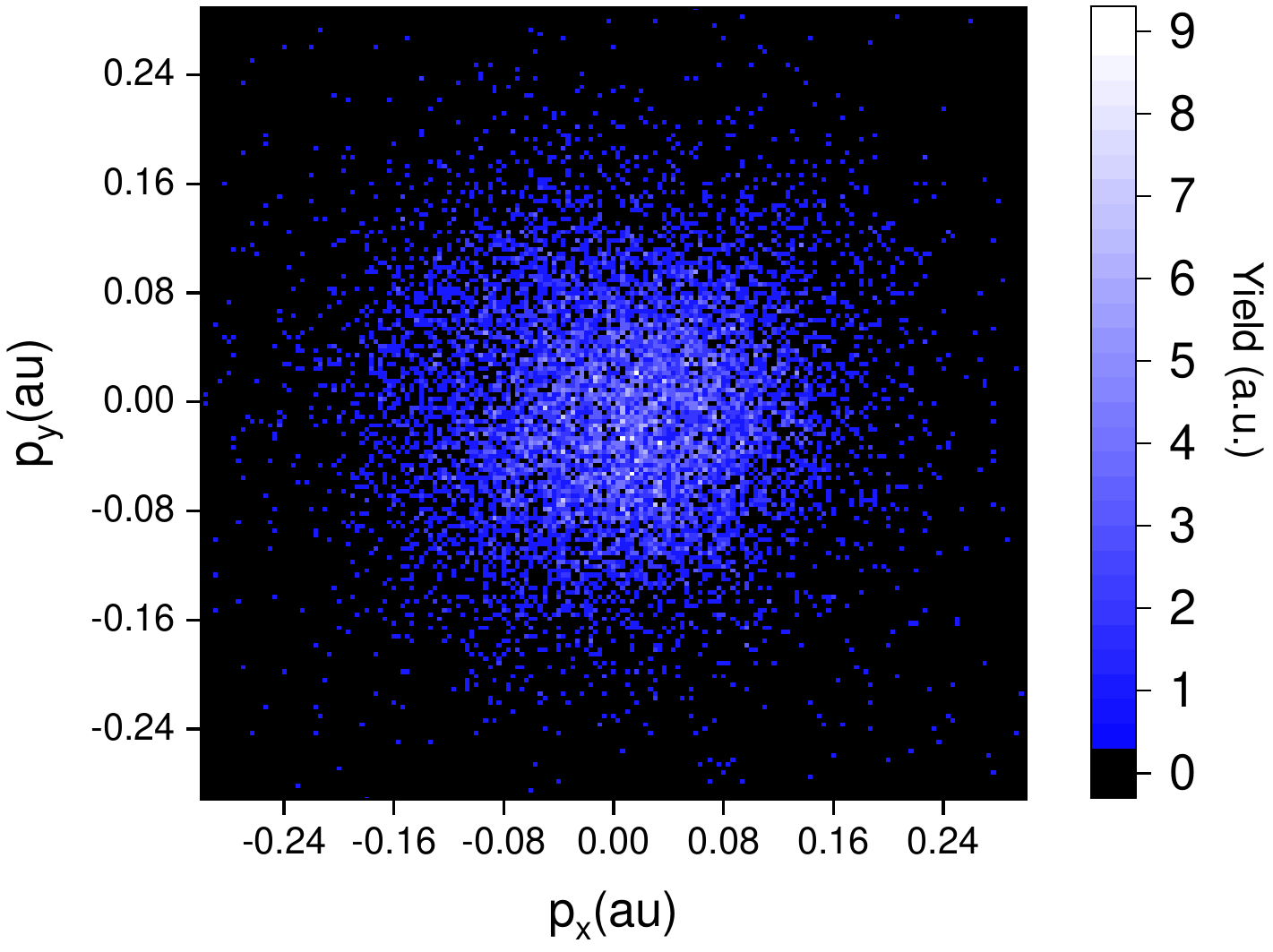}
\caption{\textbf{VMI of ICD electrons}:  VMI 
showing isotropic emission of ICD electrons from Ar atoms 
and pyridine molecules, where the imaging is not done in 
coincidence with any particular cation. The spectrum was recorded at a laser intensity of about 7 $\times$ 10$^5$ Wcm$^{-2}$. au: atomic units.}
\label{vmi}
\end{figure}
Velocity map image (VMI) of the electrons emitted 
in the present experiment, i.e., on irradiation of the 
pyridine-argon environment, is shown 
in Figure \ref{vmi}. The VMI shows isotropic emission of 
electrons with kinetic energies ranging between 0-0.6 eV. The photoexcitation energy is indeed 3.2 eV above the ionization energy of argon. However, a portion of the electronic excitation energy is rapidly transferred into vibrational degrees of freedom of four photoexcited pyridine molecules through intramolecular vibrational redistribution, reducing the electronic energy available for ionization. Importantly, under these conditions the ICD channel can open without any significant back-transfer of vibrational energy into electronic energy. Instead, only the excess electronic energy available above the ionization threshold of the Ar atom  at the moment of ionization is transferred to the ionizing electron. As a result, the emitted electrons carry very little surplus energy and therefore appear at low kinetic energies (0-0.6 eV). This rapid redistribution of electronic energy into IVR is fully consistent with the high-appearance-energy fragmentation channels reported in our earlier work \cite{barik2022}. Moreover, within the collective ICD domain, low-energy transfer events are highly probable. Therefore, under our experimental conditions, high-kinetic-energy electrons are not observed. In other words, since the ICD rates are higher for a low-energy transfer, ICD electrons are typically slow \cite{jabbari}.

Figure~\ref{Ar_rel} presents direct and conclusive evidence for collective energy transfer from photoexcited pyridine molecules to non-photoabsorbing Ar atoms via ICD. The relative Ar$^+$ yield reveals a clear mechanistic shift with decreasing light intensity. At higher intensities, ICD between photoexcited pyridine molecules dominates, while at lower intensities, intermolecular energy transfer from photoexcited pyridine molecules to non-excited Ar atoms becomes the prevailing pathway. Importantly, the collision-driven ICD channel dominates across the entire intensity range investigated, rendering MPI of argon negligible.

The monotonic increase in Ar$^+$ yield with decreasing intensity reflects the progressive suppression of pyridine-pyridine ICD, in favor of the ICD mediated by argon. This interpretation is independently validated by the observed enhancement in Ar$^+$ yield with increasing argon density at fixed intensity, which confirms that energy from photoexcited pyridine ensembles is efficiently pooled and transferred to ionize a non-photoabsorbing species.

These findings are consistent with the proposed molecular mechanism: at high laser intensity, the reaction rate scales with intensity due to ICD between multiple excited pyridine molecules alone, without the involvement of argon. At such high laser intensities, ICD processes within pyridine gas alone, without the involvement of Ar, also become more competitive  due to enhanced excitation densities. Under these conditions, the increased photon density leads to a higher population of photoexcited pyridine molecules, making ICD among three excited pyridine molecules increasingly probable. This can occur either via a genuinely concerted three-body interaction or through a sequential two-body mechanism, in which an excited dimer interacts with an additional excited monomer, corresponding to mechanisms (1) and (2) discussed in Fig. 4 of Barik et al. \cite{barik2022}. At lower excitation intensities, however, the reduced density of excited pyridine molecules renders these channels less effective, while collision-assisted ICD involving argon becomes operative. At such low laser intensities, the reaction becomes density-limited and governed by two-body collisions between a fast-moving Ar atom and transiently formed non-interacting dimer-dimer assemblies. In this regime, the number of such excited units is linearly dependent on the excitation rate, resulting in a net linear scaling of Ar$^+$ yield with intensity. 

\color{black}

\begin{figure} [H]
\centering
\includegraphics[scale=1]{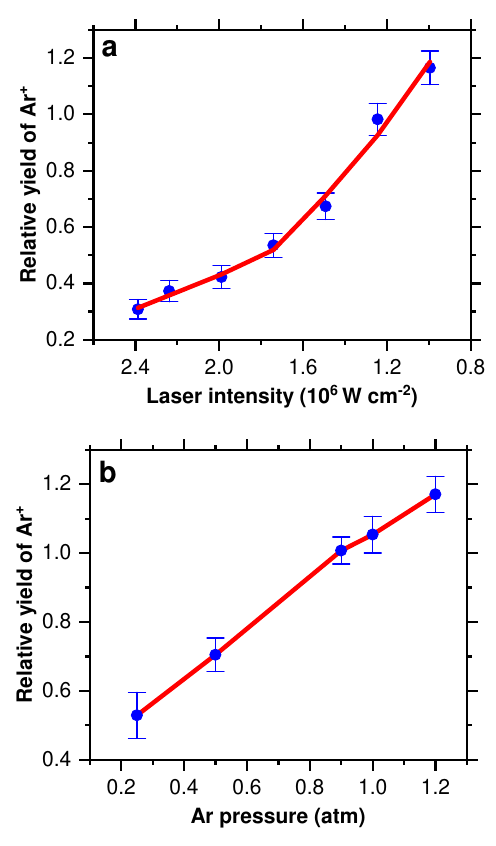}

\caption{\textbf{Relative yield of Ar$^+$:} The ratio of 
the Ar$^+$ yield to the total yield of all the other cations 
originating from 
photoexcited pyridine monomers is displayed. \textbf{(a)} The laser intensity is varied 
and the argon pressure is fixed at 0.5 atm. \textbf{(b)} The 
argon pressure is varied for a fixed laser intensity of 
$1.49 \times 10^6$~Wcm$^{-2}$. For the details on the partial pressure of pyridine monomers, see the Supplemental Material~\cite{NOTES}. The systematic increase in relative Ar$^+$ yield with decreasing laser intensity or increasing argon pressure constitutes unambiguous experimental evidence of concerted energy transfer from photoexcited pyridine molecules to non-photoabsorbing Ar atoms via the ICD mechanism.} 
\label{Ar_rel}
\end{figure}

Our study demonstrates that argon ionization in a pyridine-argon mixture arises exclusively from a concerted, nonlocal energy-transfer mechanism in which multiple $\pi^*$-excited pyridine molecules collectively channel their excitation energy to a non-photoabsorbing argon atom. This first experimental observation, which by definition constitutes collective intermolecular Coulombic decay, represents a major advance in ICD science.
This mechanism, validated through collision-free skimmed beam experiments, linear laser-intensity scaling, isotropic electron emission, and density-dependent behavior, represents a fundamentally new mode of energy redistribution. While this first report firmly establishes the existence of the process and discusses a plausible collisional scenario for its occurrence, the results reported here provide a compelling starting point for future experimental and theoretical investigations aimed at elucidating microscopic details, assessing the generality of the mechanism, and exploring its broader implications and potential applications. As a first indication of this broader relevance, the generality of the ICD mechanism is further supported by complementary experiments on pyridine-N$_2$ and quinoline-Ar mixtures at 266 nm, where ionization of the partner species is observed (see \cite{NOTES} for details). The associative interactions in the photoexcited state is a critical requirement for this mechanism. For e.g., we could not observe this
concerted-ICD ionization of Ar when we replaced pyridine with molecules like styrene. Our
theoretical calculations also revealed lack of associative interactions in photoexcited styrene molecules. The present results have stimulated theoretical studies on collective emission of virtual photons by multiple excited clusters \cite{virtual}.  More generally, these results are relevant to multiple $\pi$-molecular chromophores, as found in molecular light-harvesting systems and biomolecular systems such as DNA, where simultaneous excitation can generate excess electronic energy that must be redistributed nonlocally, either to harvest light efficiently or to prevent photodamage. Unlike these molecular light-harvesting systems, which rely on resonance with charge-separated states, the ICD pathway is driven solely by the excess electronic energy available in the system. The donor states can extend beyond $\pi$-$\pi^*$ excitations to include $n$-$\pi^*$ and Rydberg states, positioning ICD as a general framework for initiating photoreactions at distant reaction centers. Beyond energy upconversion, this mechanism may also underlie the protective role of $\pi$-chromophores in safely dissipating excess excitation energy nonlocally, thereby safeguarding biomolecular systems under UV exposure.
\section*{Acknowledgments}
 This work was supported by the Department 
of Science and Technology through project nos: 
EMR/2016/005247 (G.A.) and CRG/2022/003516 (G.A.).
\section*{Data Availability}
The data that support the findings of this article are openly available \cite{data}.
\newpage
\section*{\textbf{SUPPLEMENTAL MATERIAL}}
\section{Kinetic energy spectrum of the emitted electrons.}

\begin{figure}[!h]
	\centering
	\includegraphics[scale=0.4]{FIGURE4.pdf}
	\includegraphics[scale=0.4]{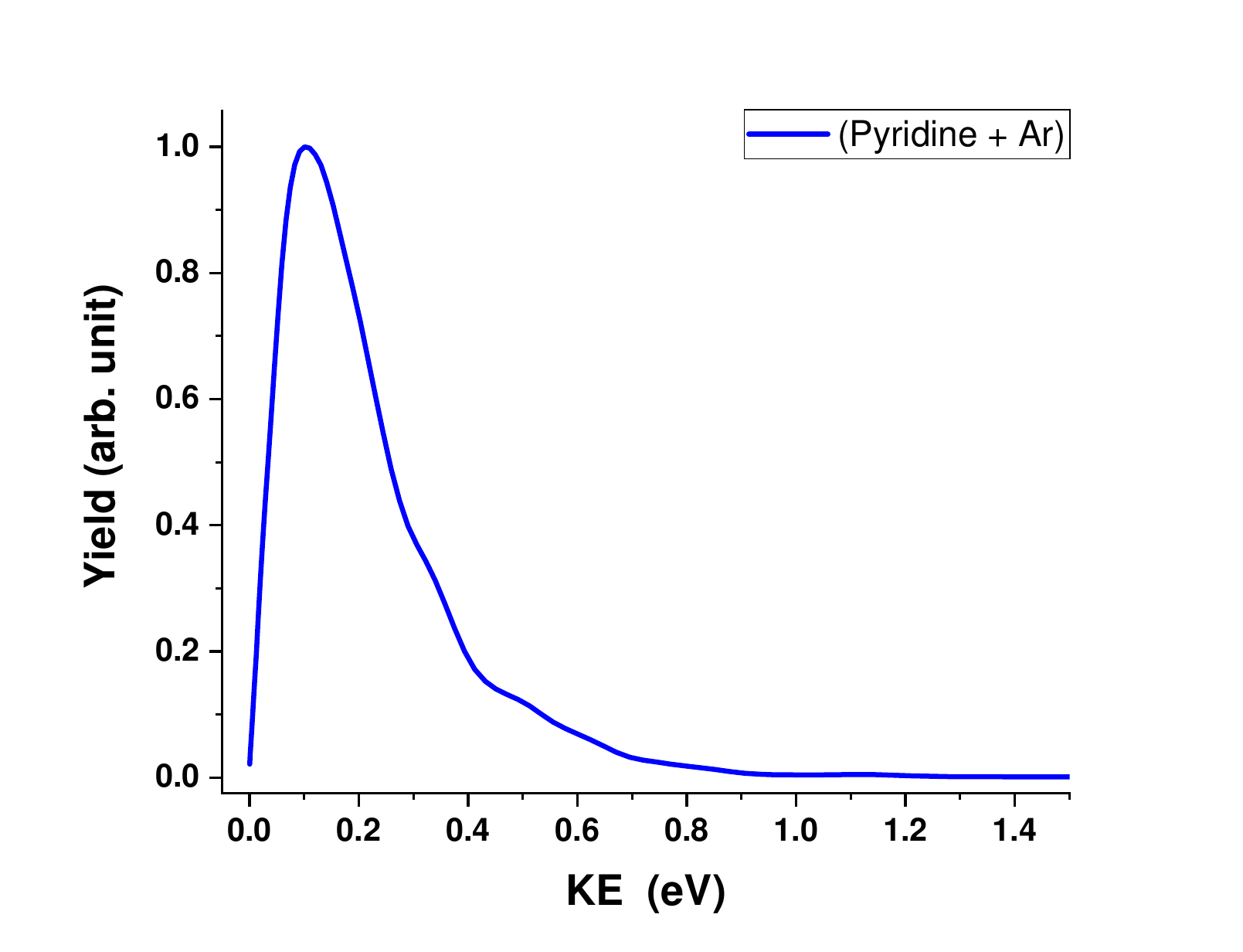}
	\caption{Velocity Map Imaging (VMI) of electrons: The top panel displays the VMI of the electrons emitted and the bottom panel displays the kinetic energy of the released electrons. au, atomic units. a.u., arbitrary units.}
	\label{vmi}
\end{figure}

Figure \ref{vmi} shows the velocity map image of the ICD electrons with low kinetic energies.  In most of the ICD experiments reported so far, excitation of one sub-unit of a dimer is the first step. Then an ultra-fast energy transfer from the excited sub-unit to the other sub-unit of the dimer ionizes the latter. In such cases, the maximum kinetic energy of the ejected electrons will be the difference between the excitation energy in the dimer and the ionization energy.   However, in the present experiment, since the initial excitation is in unbound molecular monomers, we expect slow ICD electrons, as discussed below.  On photoexcitation of pyridine monomers, Intramolecular vibrational redistribution (IVR) leads to transfer of the excess electronic energy into the nuclear modes. When ICD occurs \textit{en route} to the association of these excited monomers, only a small fraction of energy in the nuclear modes becomes available for ICD ionization leading to slow electrons. This is because ICD occurs faster than the relaxation of energy from the nuclear modes back to the electronic mode. Since IVR is the characteristic of the photoabsorbing $\pi$-molecules, the range of kinetic energy for the ICD electrons should not differ much for the ICD ionization of pyridine molecules \cite{barik2022} and Ar atoms. Further, the observed isotropic emission of the ICD electrons is characteristic of ICD ionization.

\section{Collision-free and Collision-prone experimental arrangements}

\noindent \textit{Experimental set up:} Collective energy transfer between photoexcited pyridine molecules and argon requires them to be in close proximity. At close proximity, $\pi^*$-excited pyridine molecules can undergo associative interactions while simultaneously allowing overlap with the electron cloud of a nearby argon atom. Hence, a collision-prone interaction region is necessary for facilitating the same. High density of pyridine molecules and argon atoms at the interaction region is another critical requirement for increasing the ICD events to a detectable level. A pulsed supersonic-jet expansion was chosen to achieve a high density of the ICD participants at the interaction spot where laser collides with the molecules. This interaction spot is midway between the first two electrodes of the time-of-flight (TOF) mass spectrometer. To enhance the collisions between molecules thereby bringing the molecules in close proximity, the supersonic expansion was directed between the first two TOF electrodes, which were 10 cm in diameter and 2 cm apart. Such an expansion results in shock waves and intense collisions between the two electrodes. The supersonic expansion from the nozzle results in angular divergence of the molecular beam, allowing a major fraction
of the molecules to enter the narrow 2 cm-wide gap between the two TOF electrodes positioned immediately in front of the nozzle. Within this confined geometry, the angularly diverging beam undergoes multiple scattering events from the electrode walls, leading to re-injection of molecules into the central axis of the beam path en route to the interaction region, situated approximately 5 cm downstream. This repeated
scattering establishes a quasi-crossed molecular beam configuration characterized by multiple intersecting trajectories. As demonstrated in our previous electron-impact ionization studies [21], such collisional conditions facilitate efficient cluster dissociation into monomers and enable the spatial proximity between monomers necessary for ICD to occur. 

\begin{figure}[!h]
	\centering
	\includegraphics[scale=0.35]{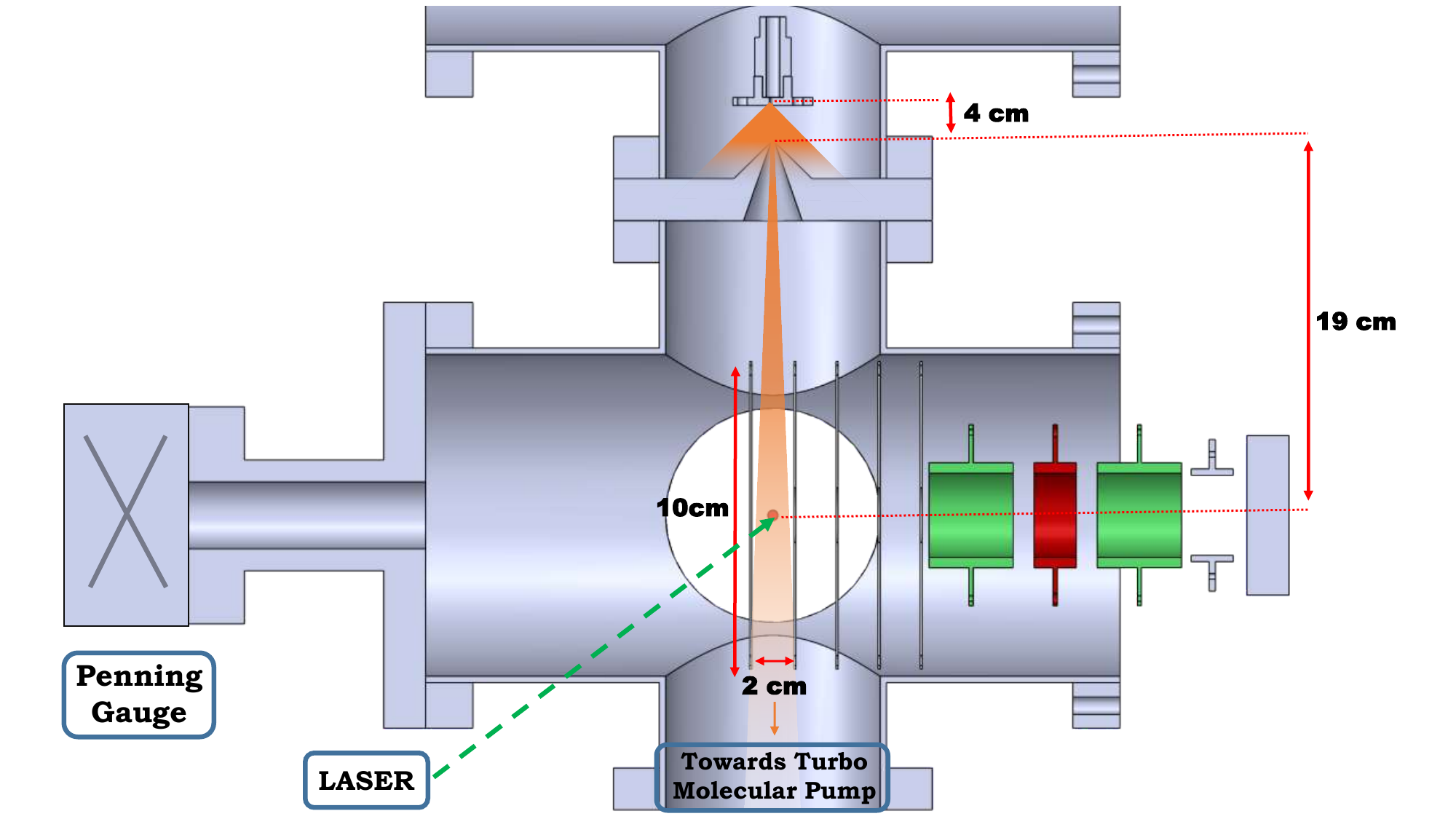}
	\vskip 10pt 
	\includegraphics[scale=0.35]{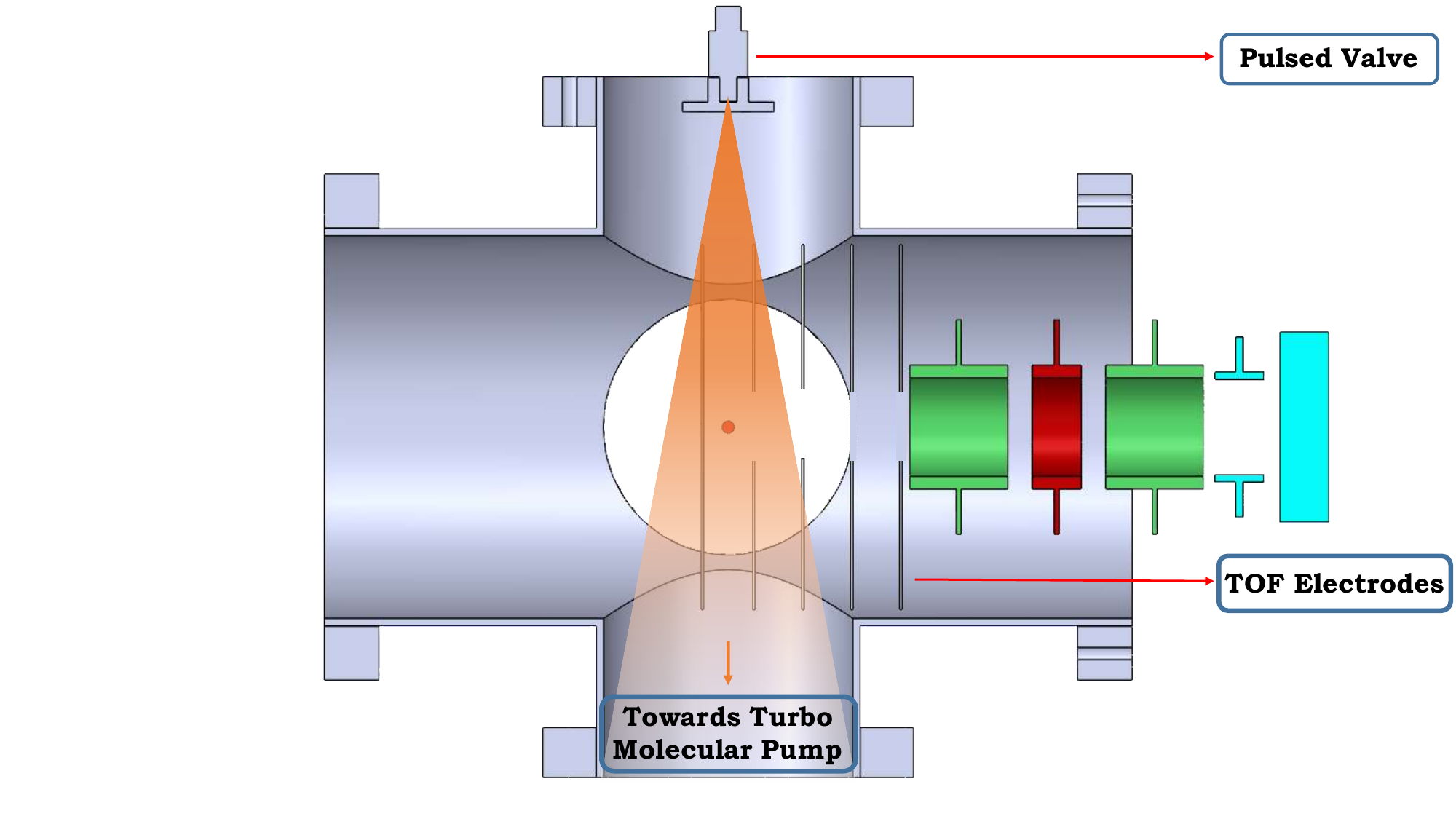}
	
	\caption{\textbf{Schematic illustration of TOF mass spectra of cations under skimmed (top panel) and unskimmed (bottom panel) beam conditions.} In the skimmed configuration, the pyridine-Ar mixture enters the interaction region between the first two TOF electrodes under essentially collision-free conditions, whereas the unskimmed beam leads to frequent intermolecular collisions.}
	\label{massspec}
\end{figure}

\section{Additional details on the experiment}

\noindent \textit{Electron measurements:} The VMI electrodes were plated with gold, whose workfunction is above the photon energy (4.661 eV). 

\noindent \textit{Ion measurements:} The photon energy is above the workfunction of the stainless steel TOFMS electrodes ($\approx$ 4.4 eV). In order to avoid the acceleration of the photoelectrons, if any,  produced from the surfaces of the TOFMS electrodes, we had applied the TOFMS pulsed electric-field with a finite time-delay \textit{after} the laser-pulse. This delay was about 1 microsecond, which is sufficient to get rid of the photoelectrons ($\approx$0.26 eV) from the volume contained by the TOFMS electrodes. We did not apply an electric-field exclusively to sweep photoelectrons produced, if any. We had also performed SIMION simulations, which show that even if photoelectrons were produced from the TOF electrode surfaces and are accelerated, their path would result in the formation of Ar$^+$ \textit{much} away from the central volume of the first two TOF plates. In such a situation we see that the field is unable to manoeuvre the ions to the detector, which is located 1 meter away from the interaction region. TOF electric field is optimized to collect ions produced at the overlap of laser-gas beam region. Ions produced at random locations will also result in a very broad TOF peak, which is also not seen for any of our ions. Thus, with great care we have ensured the absence of any role of photoelectrons.

\begin{figure}[H]
	\centering
	\includegraphics[scale=0.4]{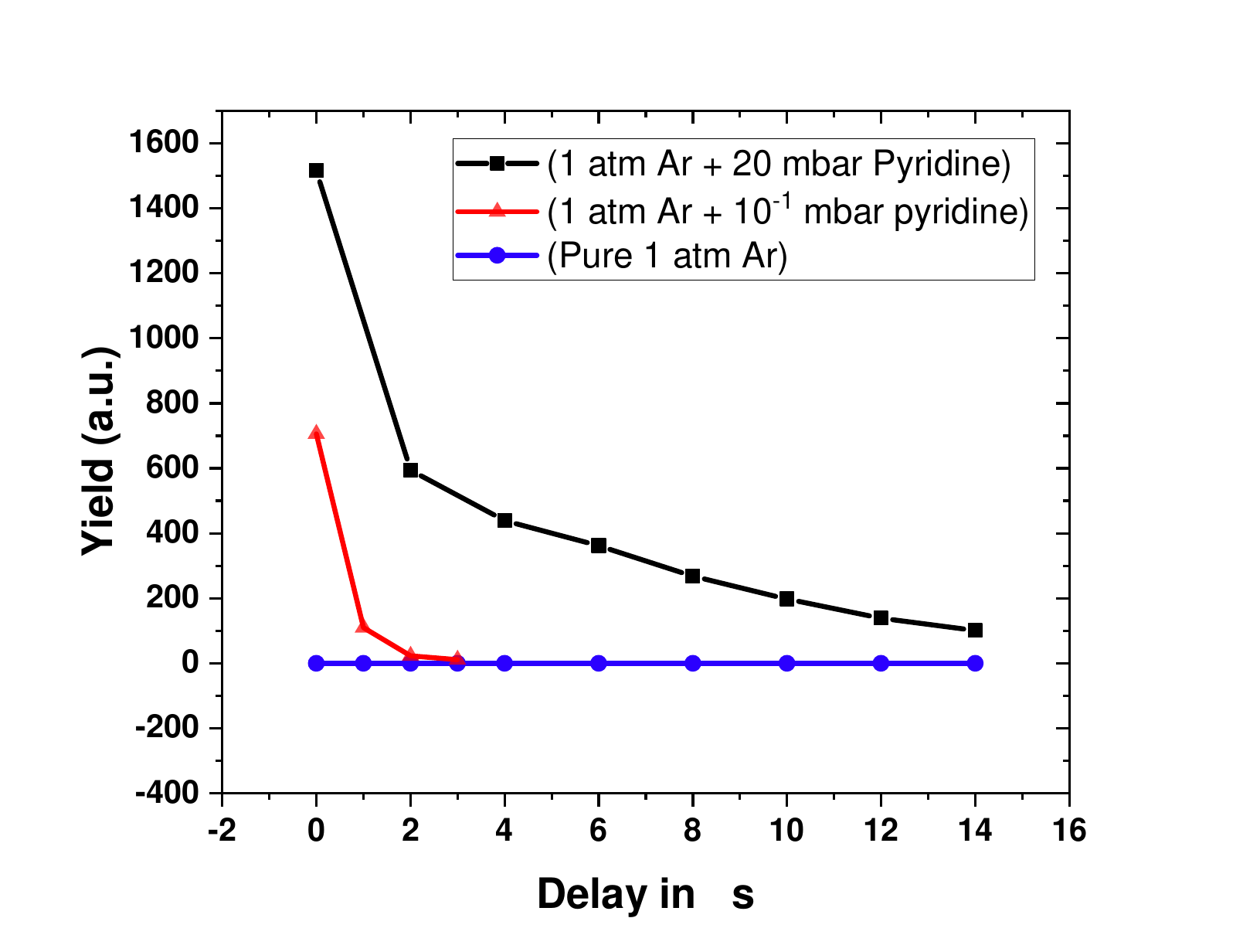}
	\caption{The yield of Ar$^+$ cations as a function of the delay between the laser pulse and the TOFMS electric-field pulse. The laser energy was maintained the same for these data points.}
	\label{Ar+}
\end{figure}

Thus Ar$^+$ were not formed on irradiation of \textit{pure-Ar gas alone}, as shown in the Figure \ref{Ar+} (blue points).  We observe Ar$^+$ only when \textit{pyridine-Ar gas mixture} is irradiated and most importantly, the Ar$^+$ yield depends on the density of the pyridine gas, as seen by comparing the black and red data points in the Figure \ref{Ar+}. The latter clearly asserts that the ionization of Ar is due to concerted energy-transfer from photoexcited pyridine molecules. Note that the yield in Figure \ref{Ar+} decreases with increasing time-delay. This is due to the degrading collection efficiency as the initial position of the ions deviate from the volume centered between the first two TOFMS electrodes.

\clearpage

\pagebreak

\section{Absence of electron-impact ionization of Argon in the present experiments}

\subsection{Main experiment}
Below, we describe the experiments that rule out electron-impact ionization of argon in our study.

\begin{itemize}
	\item[] Figure \ref{tof} shows the schematic of our time-of-flight spectrometer (TOF) along with the voltages applied. The distance between adjacent electrodes = 2 cm.
	\begin{figure} [!h]
		\centering
		\includegraphics[scale=0.5]{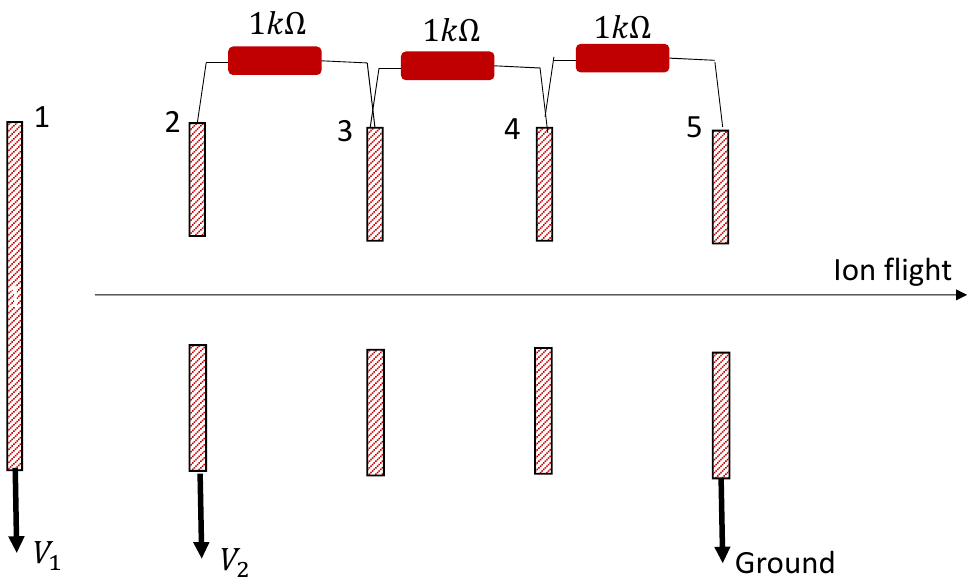}
		\includegraphics[scale=0.5]{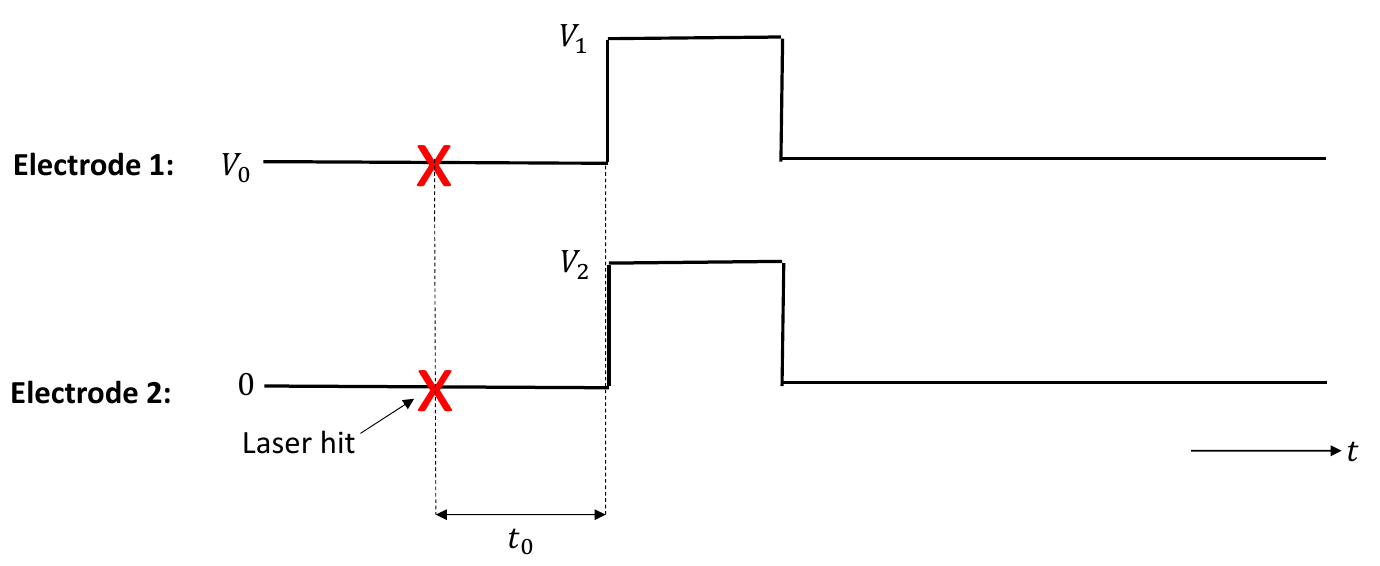}
		\caption{ \textit{Electrode 1}: A pulsed voltage $V_1$ is applied over a DC base voltage $V_0$. \textit{Electrode 2}: A pulsed voltage $V_2$ is applied w.r.t. to the ground. The ions and electrons formed between the first two electrodes experience a DC electric field of 50$V_0$ V/m for $t_0$ microseconds, before the application of the high voltage pulses that drive the cations towards the detector. }
		\label{tof}
	\end{figure}
	
	\item[] For a DC potential difference of $V_0$ between the first two electrodes, the acceleration for an electron is  \begin{equation*}
	a=\cfrac{Eq}{m_e} =  \cfrac{50V_0 * q}{m_e} = 8.8 V_0\times 10^{12} ~m/s^2
	\end{equation*} 
	\textbf{When $V_0= 0.5 V$:}\\
	\item[] 
	
	(a) The time taken by an electron produced at the center of the interaction region, with zero KE, to hit the first TOF electrode is
	\begin{equation*}
	t = \sqrt{\cfrac{2\times 0.01 m}{4.4 \times 10^{12} m/s^2}} = 67 ~ns
	\end{equation*}
	
	\item[]  (b) An electron produced with 0.2 eV and with an initial velocity directed away from the first TOF electrode will travel 8 mm before it turns around. The total time for it to hit the first TOF electrode is
	\begin{equation*}
	t = \left(\cfrac{1}{a} \times \sqrt{\cfrac{2\times 0.2 ~eV}{m_e}}\right) + \sqrt{\cfrac{2\times 18 ~mm}{a}} \approx 150 ~ns 
	\end{equation*}

	\item[] Therefore for $V_0= 0.5 V$ and a delay of $1 ~\mu s$  between the laser-hit and the high-voltage extraction, there won't be any electrons present in the interaction region when the high-voltage pulse arrives!

	\begin{figure} [!h]
		\centering
		\includegraphics[scale=0.3]{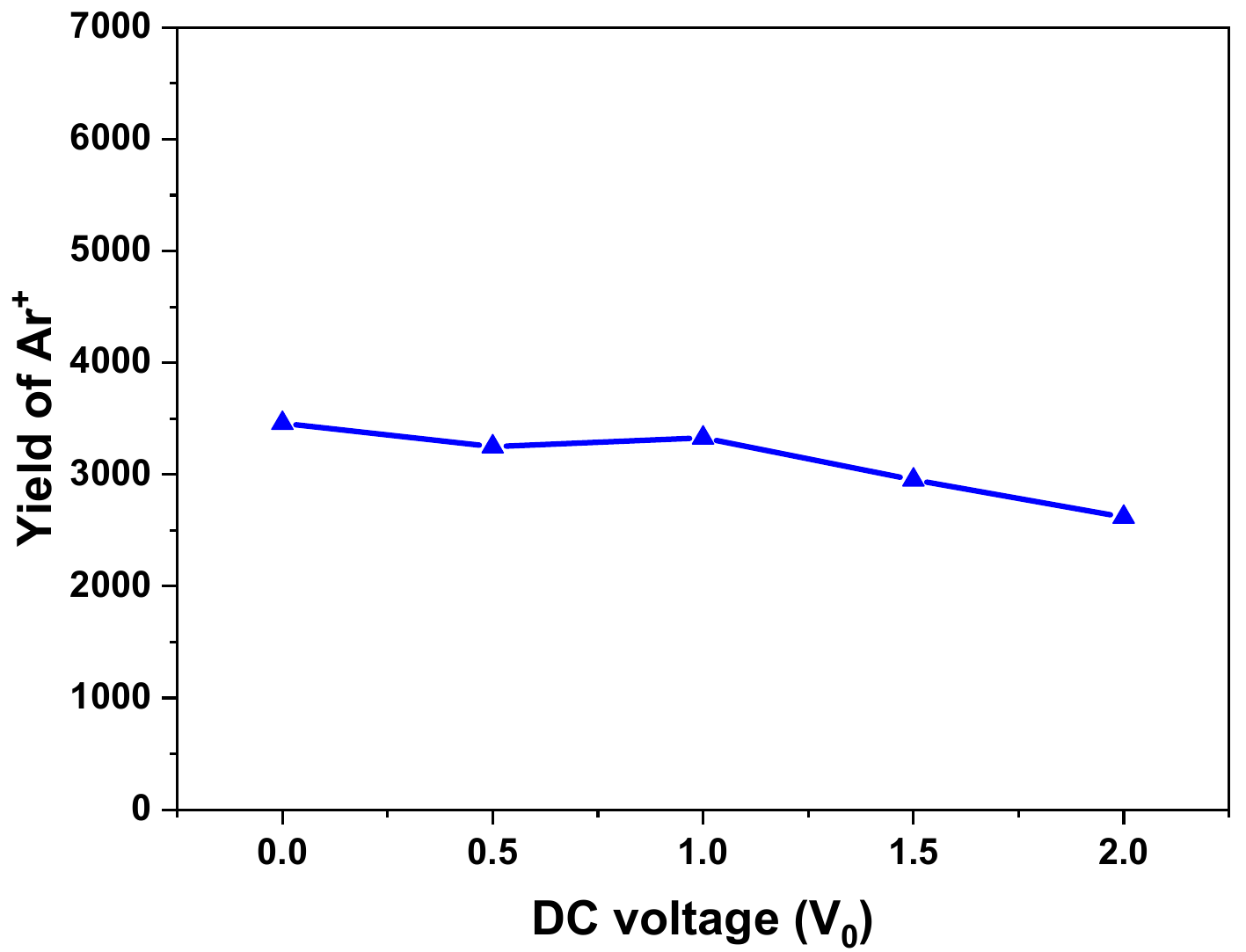}
		\caption{ The yield of Ar$^+$ cations as a function of $V_0$, the DC voltage that sweeps the electrons.}
		\label{dc}
	\end{figure}

	\item[] Figure \ref{dc} shows that the yield of Ar$^+$ is nearly the same with and without the application of the electron sweeping voltage $V_0$. For different $V_0$ values, the initial positions and energies of the Ar$^+$ cations prior to the application of the high-voltage pulse differ, resulting in slight variation in the efficiency of their collection at the detector, which is situated one meter away from the interaction point. This reflects in the minor variation in the Ar$^+$ yield as $V_0$ changes.  Thus, there is no contribution of ICD electrons in ionizing the Argon atoms. 
	
	\item[] \textbf{Earth’s magnetic field:} Since the above experiments reveal the absence of any electron during the arrival of the high-voltage pulse, the Earth's magnetic field does not also have a role here.
	
\end{itemize}

\subsection{Additional experiment}

We also performed an additional experiment, wherein $V_0=0$ but the potential difference between the first two electrodes was maintained at 10 Volts! $V_1 = 402V$ and $V_2=392 V$. Thus the maximum energy gained by any electron is 10 eV, which is insufficient to ionize Argon (I.P. = 15.6 eV).

\begin{figure} [!h]
	\centering
	\includegraphics[scale=0.42]{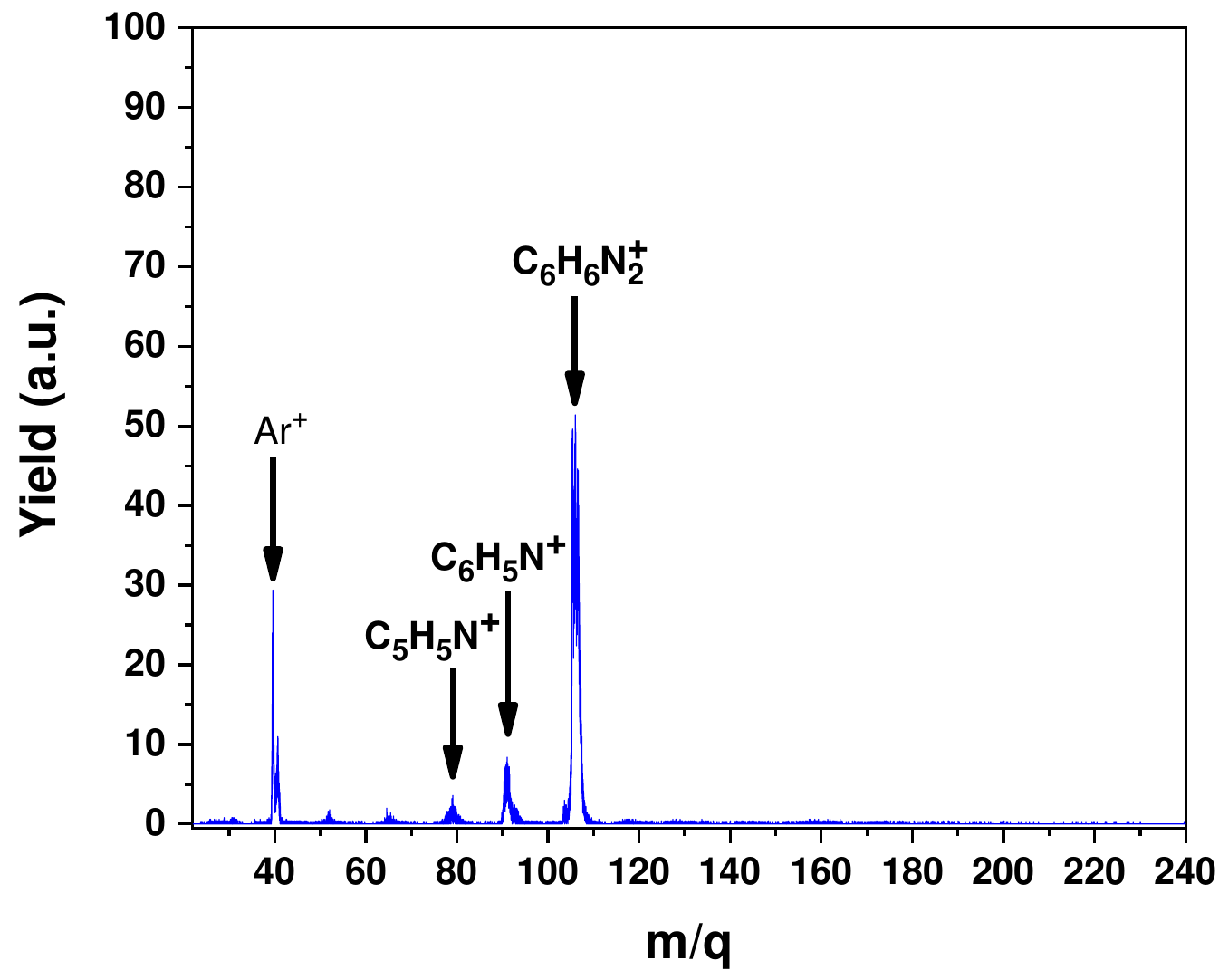}
	\caption{Mass spectra obtained by applying $V_1 = 402V$ and $V_2=392 V$. Laser and gas-beam conditions are the same. No electron sweeping potential is applied. } 
	\label{10}
\end{figure}

Figure \ref{10} shows that the mass spectra thus obtained is the same as the mass spectra obtained in our original experiments (Figure 1 of the main manuscript). 
Thus, there is no electron-impact ionization of Argon in our experiments. 
\begin{itemize}

	\item[] (a) In fact, the kinetic energy of the ICD electrons from pyridine peaks at 80 meV, corresponding to a velocity of 1.7 $\times 10^5$ m/s. Hence the ICD  electrons will move away from the interaction region (2 cm wide) in 0.1 $\mu s$. Therefore these electrons will not be present when the high-voltage pulse arrives 1 $\mu s$ post laser-hit, as done in our original experiments. 
	
	\item[] (b) Moreover, collisional ionization of Ar by the ICD electrons from pyridine, being a secondary process, should result in a very low yield of Ar$^+$ relative to the pyridine cations. On the contrary, as shown in Fig 4 of the main manuscript, with \textit{decreasing} laser intensity, the relative yield [Ar$^+$] / [pyridine cations] tends to 1!
	
	\item[] Points (a), (b), the results of the weak DC-electric field experiment and the additional experiment described above, together with the observed variation of the relative yield [Ar$^+$] / [pyridine cations] with the laser intensity, unambiguously assert the concerted ICD ionization of Ar in our experiments.
\end{itemize}
\clearpage
\pagebreak

\section{Ion-TOF spectra with no gas and with only Argon gas}

We also present the ion-TOF spectra for (a) no gas (background) and (b) only argon gas.

There is no Ar$^+$ signal seen, when only pure argon gas is irradiated with 266nm. These two spectra were taken even without a delay in the extraction pulse \textit{w.r.t} to the laser pulse. These spectra also assert that there is no formation of  Ar$^+$ due to impact with accelerated photoelectrons. These data were taken with the laser intensity and data collection time same as that for Fig 1 of the manuscript and hence the yields could be compared with each other. Only a very weak background ion signal is seen.

\begin{figure} [!h]
	\centering
	\includegraphics[scale=0.35]{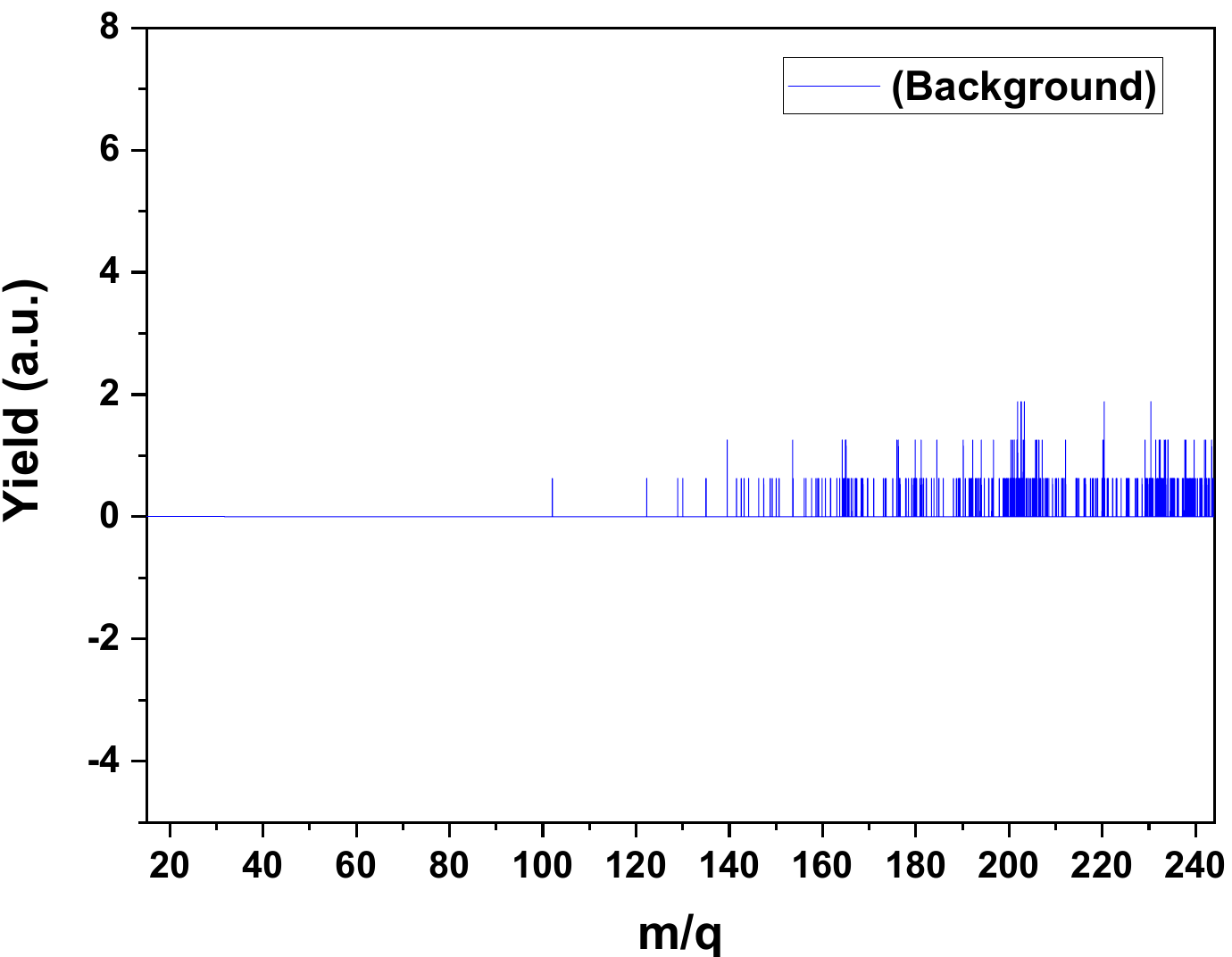}
	\includegraphics[scale=0.35]{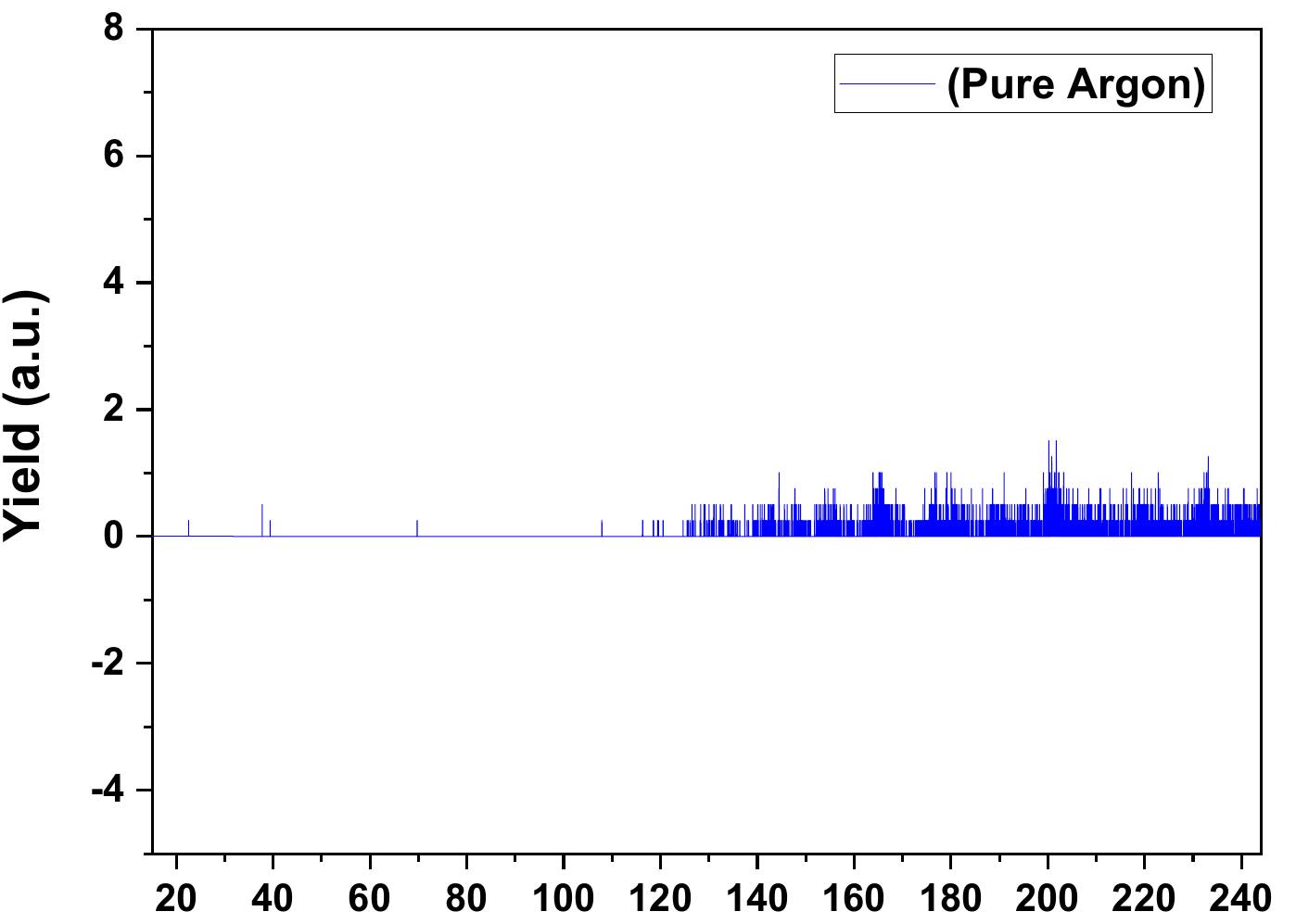}
	\caption{Ion TOF spectra for the background and pure Argon gas are presented. These data were taken with the laser intensity and data collection time same as that for Fig 1 of the manuscript.}
	\label{fig:enter-label}
\end{figure}

\section{Electron spectra with no gas and with only Argon gas}

\begin{figure}[!h]
	\centering
	\includegraphics[scale=0.35]{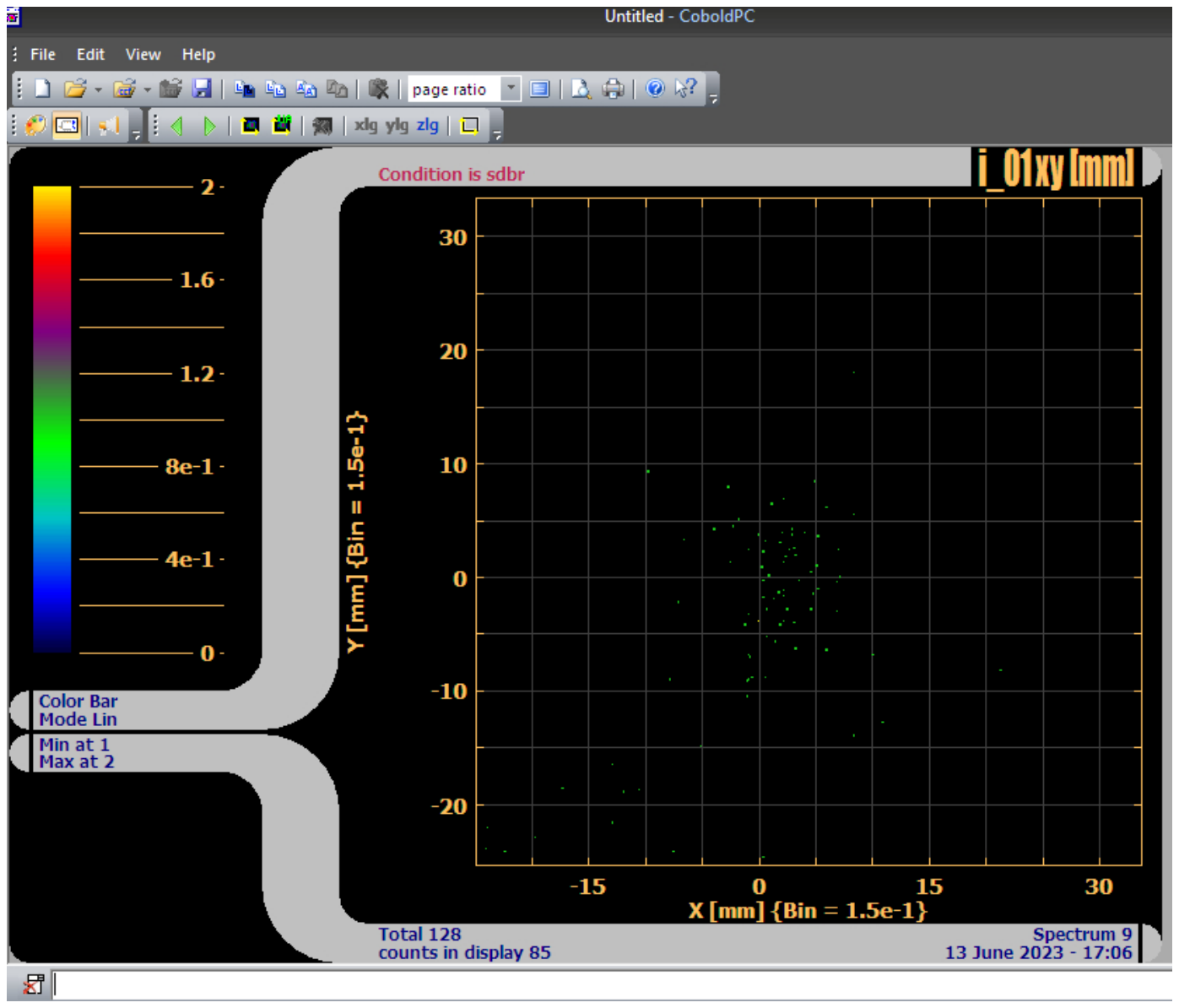}
	\caption{Electron VMI with no gas. The data is taken for 10000 laser shots with intensity same as employed for Fig 2. of our manuscript. This weak structureless spectra has count equal to 85.}
	\label{ng}
\end{figure}

\begin{figure}[h]
	\centering
	\includegraphics[scale=0.35]{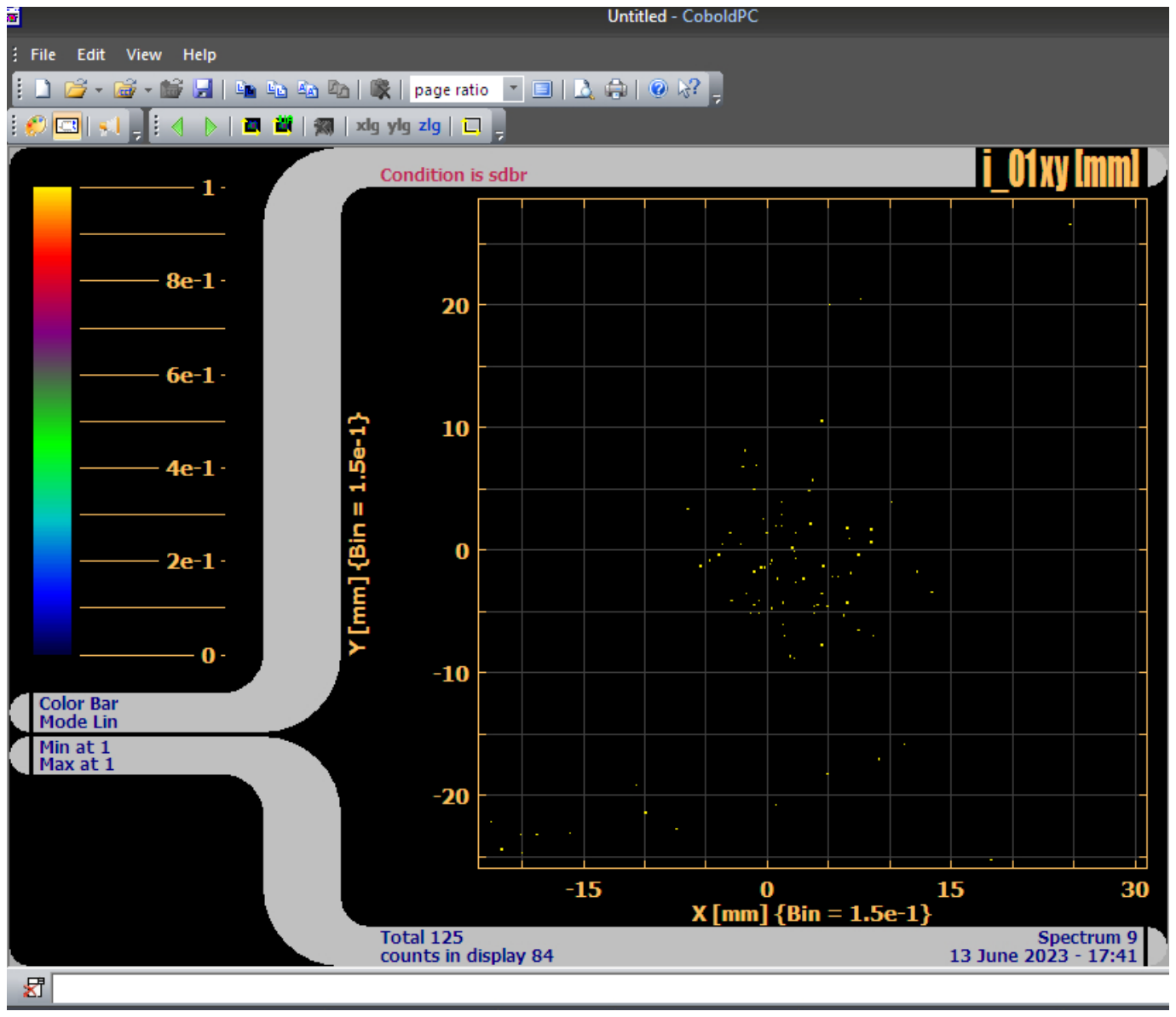}
	\caption{ Electron VMI with only argon gas. The data is taken for 10000 laser shots with intensity same as employed for Fig 2. of our manuscript. This weak structureless spectra has count equal to 84.}
	\label{g}
\end{figure}

The spectra Fig. \ref{ng} and Fig. \ref{g} are the electron spectra obtained for (a) no gas and (b) argon only, respectively. The data were taken for 10000 laser shots with the same laser intensity as employed in Fig 2 of the manuscript. A structureless weak background electron signal is seen for both the spectra with equal counts! This unambigoussly rules out Ar$^+$ signal when only Argon is irradiated with 266nm.

\section{Rate of Argon ionization}

\subsection*{Density of neutral pyridine molecules}
The density of neutral pyridine molecules at 20 mbar = $4.8 \times 10^{17}$ cm$^{-3}$.\\
Considering a reduction of density downstream at the interaction region to be by a factor $p$, the density of neutral pyridine molecules at the interaction spot,
$$n= \left(4.8 \times 10^{17} \times p\right) cm^{-3}$$

\subsection*{Photoexcitation}
Photon flux per pulse in a 7 mJ pulse of 266 nm beam (6mm diameter),
$$\Phi= 3 \times 10^{16}cm^{-2}$$
The number density of photoexcited pyridine,
$$n^*= n \times \Phi \times \sigma =  4.8 \times 10^{17} \times p \times 3 \times 10^{16} \times 19.3 \times 10^{-19} = \left(2.77 \times 10^{16} \times p \right) ~cm^{-3}$$

\section*{Estimation of Bimolecular Rate Constant}

The bimolecular rate constant for dimer formation:

\[
k = \langle \sigma_a v \rangle
\]
where $\sigma_a$ is the cross-section for association.


\begin{itemize}
	\item \textbf{Effective collisional cross-section} is estimated using hard-sphere model based on a molecular diameter of $\approx 7.5 \AA $   (in the excited state):
	
	\[
	\sigma_h = \pi d^2 \approx \pi (7.5 \times 10^{-8}~\text{cm})^2 = 1.77 \times 10^{-14}~\text{cm}^2
	\]
	If $q$ is the correction to $\sigma_h$ that yields $\sigma_a$ then,
	$$\sigma_a=\sigma_h \times q$$
	\item \textbf{Relative thermal velocity} at $T = 300~\text{K}$ is:
	\[
	\langle v \rangle = \sqrt{\frac{8 k_B T}{\pi \mu}} \approx 2.84 \times 10^4~\text{cm/s}
	\]
	where $\mu \approx 79~\text{amu} = 1.31 \times 10^{-22}~\text{g}$ is the reduced mass of pyridine.
	
	\item \textbf{Resulting rate constant}:
	\[
	k = \sigma_a \cdot \langle v \rangle = 1.77 \times 10^{-14} \cdot \left(2.84 \times 10^4 \times q \right) \approx \left(5.03 \times 10^{-10} \times q \right)~\text{cm}^3/\text{s}
	\]
\end{itemize}

This value lies within the typical range for neutral-neutral collisions in gas phase at room temperature:

\[
\boxed{k \approx  \left(5.03 \times 10^{-10} \times q \right)~\text{cm}^3/\text{s}}
\]

\section*{Dimer Formation (Excited + Excited)}

Dimerization occurs exclusively via binary collisions between two photoexcited molecules:

\[
A^* + A^* \rightarrow (AA)^*
\]

Using the rate constant and interaction time $\Delta t \approx 1 \times 10^{-9}~\text{s}$ due to the long lifetime of the triplet state:

\[
n_D = k \cdot (n^*)^2 \cdot \Delta t = \left(5.03 \times 10^{-10} \times q \right) \times \left(2.77 \times 10^{16} \times p \right)^2 \times 1 \times 10^{-9}  = \left(13.93 \times 10^{15} \times p^2 \times q \right) ~\text{cm}^{-3}
\]

\section*{Interaction of two doubly-excited dimers and argon atom}

Since pyridine is highly diluted in argon, we could consider that an argon atom is always present in the vicinity of two interacting pyridine dimers. Hence, to find the rate of Ar$^+$ formation, we here calculate the rate of collision between two doubly excited pyridine dimers. \\
\[
(AA)^* + (AA)^* 
\]

Here, since the 'size' of a doubly excited pyridine dimer unit is much larger than a singly-excited monomer, the collision cross-section is expected to be higher than that for the collision between two singly-excited monomers. If we consider one order of magnitude increase in the cross section then $$k'= 10k$$
\[
n_{Ar^+} = k' \cdot n^*  \cdot n^* \cdot \Delta t = 10\times\left(5.03 \times  10^{-10} \times  q \right)\times  \left(13.93 \times 10^{15} \times  p^2 \times  q \right)^2  \times  1\times  10^{-9} ~\text{cm}^{-3} 
\]

$$n_{Ar^+} = 9.76\times  10^{14} \times  q^3p^4~\text{cm}^{-3}$$
\section*{Total Number of argon cations per Pulse}

Assuming a cylindrical laser-gas interaction volume of radius $r = 0.05~\text{cm}$ and length $l = 0.05~\text{cm}$:

\[
V = \pi r^2 l \approx 3.93 \times 10^{-4}~\text{cm}^3
\]
Total number of Ar$^+$ per pulse, $N$ is given by,
\[
N  = \eta (=0.1)~ n_{Ar^+} \cdot V = \left(9.76\times  10^{14} \times  q^3p^4\right) \times \left(3.93 \times 10^{-4}\right) \approx 38.35\times  10^{13}\times q^3p^4 
\]
where $\eta$ is the quantum efficiency for populating the triplet state.



\textbf{Estimate of $q$ and $p$ from our measurements:} In our experiments we observe about N= 1 to 6 counts of Ar$^+$ per laser pulse. For this count rate the range of values for $p$ and $q$ are shown below. The range of values obtained is $q = 178$ to $252$ and $p = 4.28 \times 10^{-6}$  to $6 \times 10^{-6}$. 
\begin{figure}[!h]
	\centering
	\includegraphics[scale=.5]{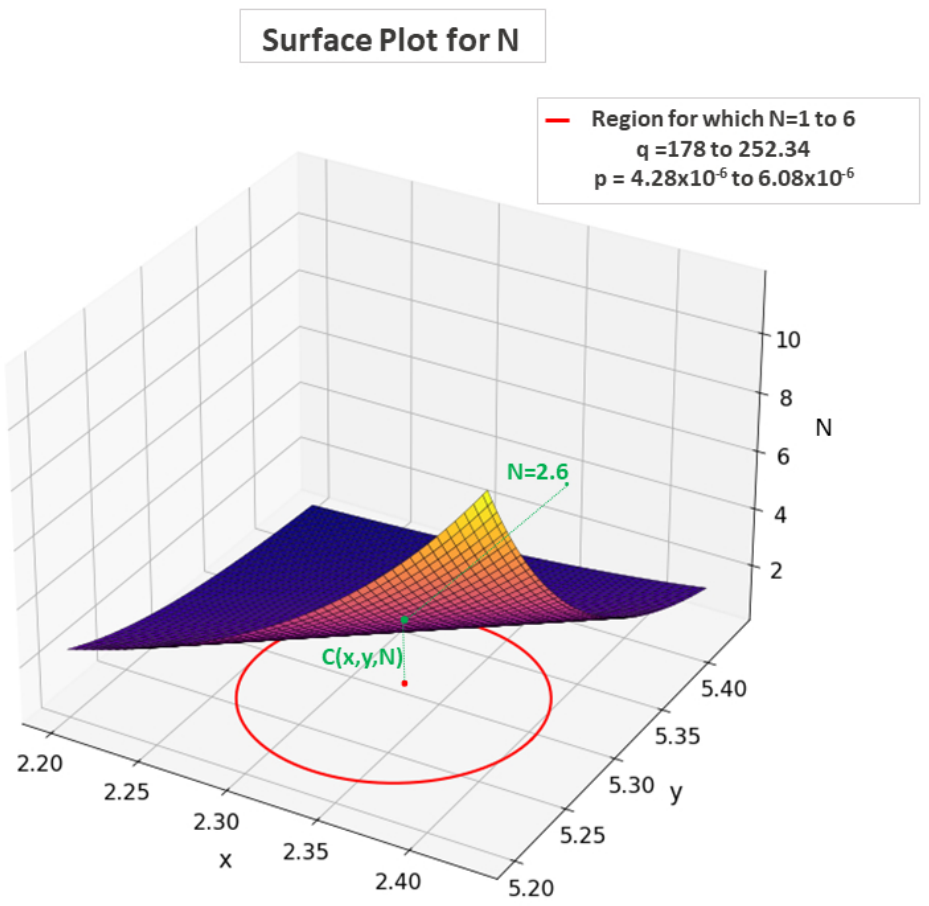}
	\caption{$q$ corresponds to $10^x$ and $p$ corresponds to $10^{-y}$. The circular area on the xy-plane, covers the range of values of x and y, and hence p and q, for the value of Ar$^+$ ranging from 1 to 6.}
	\label{fig:placeholder}
\end{figure}
These range of values for $q$ an $p$ are reasonable.
\pagebreak
\section{Irradiation of Pyridine-Nitrogen mixture and Quinoline-Ar mixture at 266 nm }

\begin{figure}[!h]
	\centering    \includegraphics[width=0.8\linewidth]{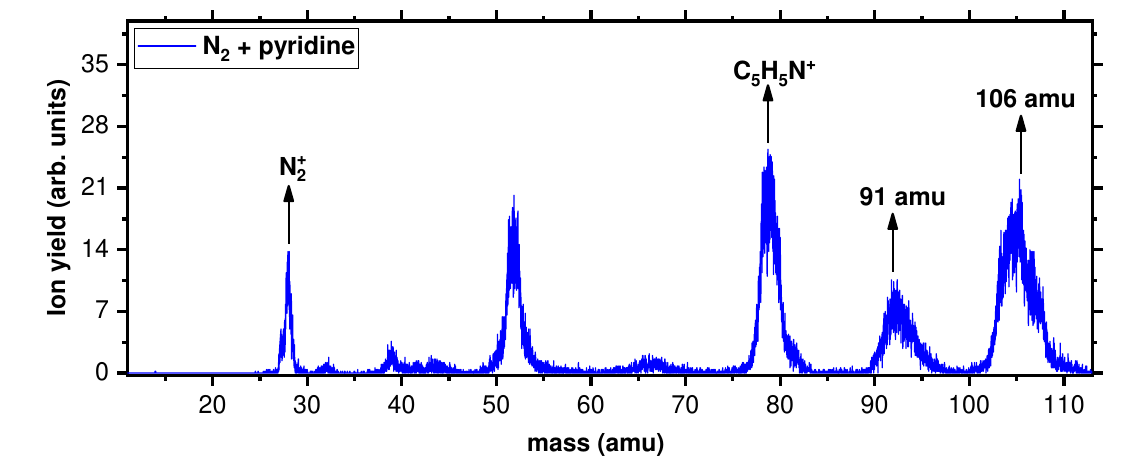}
	\caption{Spectrum obtained by the irradiation of Pyridine-nitrogen gas mixture at 266 nm. Nitrogen cations are observed to form.}
	\label{n2}
\end{figure}

\begin{figure}
	\centering
	\includegraphics[width=0.5\linewidth]{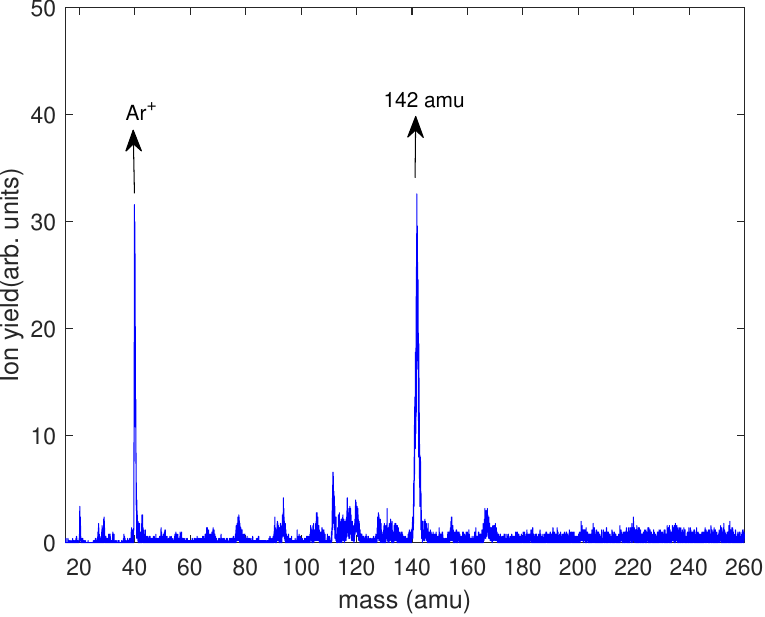}
	\caption{Spectrum obtained by irradiation of Quinoline-Argon gas mixture at 266 nm. Argon cations are observed to form.}
	\label{q+ar}
\end{figure}

Under the collision-prone experimental conditions, when we irradiated Pyridine-Nitrogen mixture at 266 nm, we could observe ionization of nitrogen. The Fig. \ref{n2} shows the spectrum for the same. Further when we irradiated Quinoline-Ar mixture we could again observe Ar ionization. The Fig. \ref{q+ar} shows the spectrum for the same. On the other hand we could not observe this ICD ionization of Ar when we replaced pyridine with molecules like styrene. Our theoretical calculations also revealed lack of associative interactions in photoexcited styrene. These results indicate that the photoabsorbing unit should be highly interacting in their excited states such that association between them occurs and that concerted energy-transfer from such photoexcited molecules to any atom/molecule could occur.   


\end{document}